\documentclass[12pt]{article}
\pdfoutput=1 
\usepackage{graphicx,amsmath}
\usepackage{units}
\usepackage{color}
\usepackage{float}

\newlength{\dinwidth}
\newlength{\dinmargin}
\def\be{\begin{equation}}   
\def\ee{\end{equation}}  
\def\bea{\begin{eqnarray}}                      
\def\eea{\end{eqnarray}}
\def\ch1{$\chi(1^+)$}
\def\lapproxeq{\lower .7ex\hbox{$\;\stackrel{\textstyle                                                    
<}{\sim}\;$}}                                                    
\def\gapproxeq{\lower .7ex\hbox{$\;\stackrel{\textstyle                                                    
>}{\sim}\;$}}       
\def\cc{$c\bar{c}~$}
\def\bb{$b\bar{b}~$}

\begin{document}

\begin{flushright}                                                    
IPPP/17/104 \\                                                    
\today \\                                                    
\end{flushright} 

\vspace*{0.5cm}

\begin{center}

{\Large \bf Open charm production and low $x$ gluons} \\
\vspace{0.5cm}
\vspace{1.0cm}

E.G. de Oliveira$^{a}$, A.D. Martin$^b$ and  M.G. Ryskin$^{b,c}$\\ 

\vspace{0.5cm}
$^a$ {\it Departamento de F\'{i}sica, CFM, Universidade Federal de Santa
Catarina, C.P. 476, CEP 88.040-900, Florian\'opolis, SC, Brazil}\\    
$^b$ {\it Institute for Particle Physics Phenomenology, University of Durham, Durham, DH1 3LE } \\
$^c$ {\it Petersburg Nuclear Physics Institute, NRC Kurchatov Institute, Gatchina, St.~Petersburg, 188300, Russia } \\ 

\begin{abstract}
We compare the rapidity, $y$, and the beam energy, $\sqrt{s}$, behaviours of the cross section of the data for $D$ meson production in the forward direction that were measured by the LHCb collaboration. We describe the observed cross sections using NLO perturbative QCD, and choose the optimal factorization scale for the LO contribution which provides the resummation of the large double logarithms. We emphasize the inconsistency observed in the $y$ and $\sqrt{s}$ behaviours of the $D$ meson cross sections. The $y$ behaviour indicates a very {\it flat} $x$ dependence of the gluon PDF in the unexplored low $x$ region around $x\sim 10^{-5}$. However, to describe the $\sqrt{s}$ dependence of the data we need a steeper gluon PDF with decreasing $x$. Moreover, an even steeper behaviour is needed to provide an extrapolation which matches on to the well known gluons found in the global PDF analyses for $x\sim 10^{-3}$. The possible role of non-perturbative effects is briefly discussed.

\end{abstract}

\end{center}
\vspace{0.5cm}

\section{Introduction}

The LHCb collaboration have published measurements of the cross section for $D$ meson~\cite{LHCbcc7,LHCbcc13,LHCbcc5} and $B$ meson~\cite{LHCbB1,LHCbB} production in the forward direction with rapidities in the region $2<y<4.5$. 
$D$ meson production is measured via the decays $D\to K\pi,~K\pi\pi$, and the data are available, differential in both rapidity and transverse momentum, at three beam energies $\sqrt{s}=$ 5, 7 and 13 TeV. The $B$ meson data, obtained via $b\to J/\psi X$ decays, are available at 7 TeV differential in both $y$ and $p_t$ \cite{LHCbB1}, but the more recent measurements via semileptonic decays at 7 and 13 TeV are presented only inclusively in $p_t$  \cite{LHCbB}.

\begin{figure} [h]
\begin{center}
\includegraphics[clip=true,trim=0.0cm .0cm 0.0cm 0.0cm,width=12.0cm]{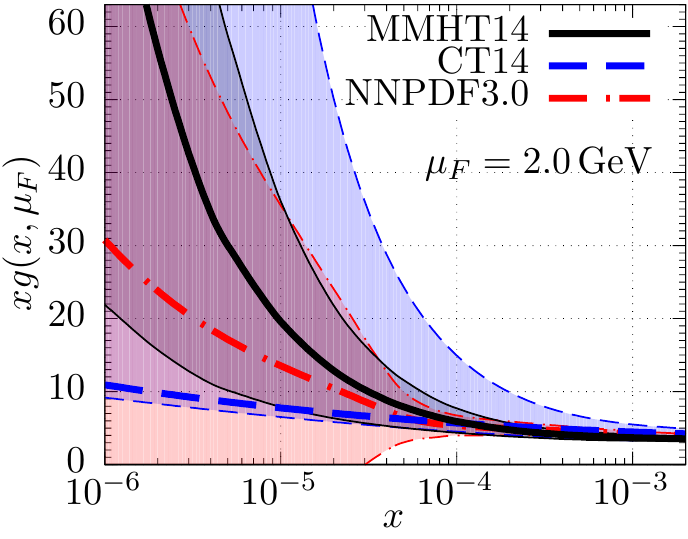}
\caption{\sf The low $x$ gluon density as given by recent NLO global parton analyses \ NNPDF3.0~\cite{Ball:2014uwa},
MMHT2014~\cite{Harland-Lang:2014zoa}, CT14~\cite{Dulat:2015mca}, calculated using the PDF interpolator LHAPDF~\cite{Buckley:2014ana}.}
\label{fig:g}
\end{center}
\end{figure}

Here, we concentrate on $D$ meson production, since we wish to probe low scale distributions which are more sensitive to the input gluon PDF.  The data can be described by the production of a 
$c\bar c$-pair followed by the fragmentation of  the $c$ quark into the $D$ meson. Moreover, gluon fusion, $gg$, is the major contributor to forward $D$ meson production. Therefore,  since the  mass of the $c$ quark, $m_c$, is not too high and that there is a large rapidity, this process allows a probe of the gluon distribution $g(x,\mu_F^2)$ at very small $x$ \cite{R1}$-$\cite{Gauld}
\be
x~\sim~ (m_T/\sqrt{s}) e^{-y}~ \sim ~10^{-5}
\ee
and at a small factorization scale $\mu_F \sim m_T$. Here $m_T=\sqrt{m^2_c +p_{t,c}^2}$, where $p_{t,c}$ is the transverse momentum of the $c$ quark and $\sqrt{s}$ is the centre-of-mass proton-proton energy.  Due to the absence of data probing the gluon in the low $x$ domain the gluon PDF is practically undetermined  in this region by the global PDF analyses. The uncertainty at a scale $\mu_F=2$ GeV is illustrated in Fig.~\ref{fig:g}.

Thus the open $c\bar{c}$ data are potentially very valuable. However a major problem is the sensitivity to the choice of factorization and renormalization scales. Note that in this low $x$ region the gluon density strongly depends on the factorization scale $\mu_F$ due to the presence of  
Double Log (DL) terms $[(\alpha_s N_C/\pi){\rm ln}(1/x){\rm ln}\mu_F^2]^n$ with a large $\ln(1/x)\sim 10$. The conventional choice of scale is $\mu_F=m_T$. Indeed, it was shown that the DL terms can be resummed into the incoming parton PDFs by choosing $\mu_F\simeq 0.85m_T$ \cite{OMR}. Though the dependence on the residual scale, $\mu_f$, becomes weaker, it is nevertheless still appreciable. 

It is known that the ratio of measured cross sections at different energies or at different rapidities is less sensitive to the value chosen for the factorization or renormalization scales~\cite{R1}$-$\cite{Gauld}. However, ratios do not determine the overall normalization of the gluon density.  On the other hand it would be valuable to fix the absolute value of the gluon density from these low $x$ data.  Even though the \cc data have much lower statistical weight in comparison to the data used in the global PDF analyses, a reliable estimate of the gluon density in the unexplored region $x\sim 10^{-5},~Q^2 \lapproxeq 10$ GeV$^2$ would therefore be of great value.  

Among the papers that have studied the impact of \cc and/or 
\bb LHCb data on the behaviour of the gluon in the low $x$ region,  two papers have direct bearing on our study.  First, \cite{R2} adds the \cc and \bb LHCb data to an ensemble of other data in a global PDF fit. The paper describes two types of fit.  The most relevant is the fit where they attempt to determine the {\it normalization} of the low $x$ gluon PDF. Since the result must be consistent with the gluons obtained from the whole set of data included in their global analysis, they find this  can only be achieved by allowing the factorization and renormalization scales for both \cc and \bb production to be free parameters.  They find that the data require the scales $Am_T$ to be 
\be
\label{eq:Acb}
A^c_F=0.66,   ~ A^b_F=0.26,  ~ A^c_R=0.44, ~ A^b_R=0.33. 
\ee
We note that $\mu_R^c=0.44m_T$ was also found in \cite{GR}. Such unnaturally low scales indicate that actually the heavy quark data needs low $x$ gluons larger than those extrapolated from the global PDF analyses \cite{Ball:2014uwa}$-$\cite{Dulat:2015mca}.  In other words the value of the cross section, $\sigma\sim\alpha^2_s(\mu_R^2)~g(x_1,\mu_F^2)..$, which was described by a low scale $\mu_R$, may be reproduced using a natural scale and a larger gluon, $g(x_1)$.

The second paper, Gauld \cite{Gauld}, concentrates on attempting to determine the low $x$ gluon from the \bb LHCb data \cite{LHCbB1,LHCbB}, and compares with results he obtains from the \cc data.  He finds a large discrepancy with these forward production data at 13 and 7 TeV, both through fits to the normalized cross sections and to various cross section ratios.  He concludes that the LHCb data are not consistent with evolution via perturbative QCD, assuming that there is {\it no} dramatic change in the input gluon behaviour in the region $x\sim 10^{-4}-10^{-3}$.

Here we proceed differently, although we, too, work at NLO. For the partonic subprocesses $a(x_1)b(x_2) \to c\bar{c}X$, which drive the forward open $c\bar{c}$ production, we take the partons from the global PDF analyses with the exception of the gluon PDF at very low $x\equiv x_1$ say.  Recall that the beam energies of the LHCb experiments sample, for low $p_t$ and large $y$, the $x$ intervals
\be
5\times 10^{-6}  \lapproxeq ~x_1 ~\lapproxeq 5\times10^{-5},~~~~~~x_2\gapproxeq 10^{-2}.  
\ee
Although we include the contributions from production by NLO $gq$ and $q\bar{q}$ fusion, we note the dominance of production by $gg$ fusion, so the assumption that only the low $x$ gluon PDF is unknown is reasonable. We therefore perform fits to the LHCb open charm data using various parametrizations of the low $x$ gluon density. 
We proceed in three stages, which may be summarised as follows.
\begin{itemize}
\item[(1)] We first explore how well a simple two-parameter form for the low $x$ gluon describes the $\sqrt{s}$ and $y$ behaviour of the data in each $p_t$ bin independently, see Section~\ref{sec:3}.
Since the gluon distribution at or near the input of DGLAP evolution is the most interesting\footnote{At large scales the gluon PDF is formed mainly by evolution depending on parton PDFs with larger $x$ at low scale.}, here we consider the LHCb data in the four $p_t$ intervals from 1 to 5 GeV.
\item[(2)] Then, more ambitiously, we fit to all the LHCb data in the interval $1<p_t<5$ GeV simultaneously using a simple two-parameter Double Log (DL) parametrization (which is known to well approximate DGLAP evolution at low $x$ throughout this $p_t$ range).
We obtain a satisfactory description of all the open charm LHCb data, namely $\chi^2=141$ for 120 $D$ meson data points, see Section~\ref{sec:4}. So far, so good. Thus, up to the sizeable uncertainty due to the choice of scales we know the behaviour of the gluon PDF around $x=10^{-5}$. The problem is that when our fit is extrapolated from $x=10^{-5}$ to $x=10^{-3}$, into the domain where the gluon is well known from the global PDF analyses we see a mismatch of the behaviour\footnote{Here the optimal scale $\mu_F=\mu_R=0.85m_T$, which resums the large DL terms, is used.}.  
\item[(3)] Indeed, the extrapolated gluon is more than a factor of two above the well known gluon PDF of the global analyses at $x=10^{-3}$. To attempt to cure this problem we re-fit the data using a lower scale than the optimal choice $\mu_F (=\mu_R)=0.85m_T$.  To be precise we take $\mu_f (=\mu_R) =0.5m_T$ and also we adjust the DL parametrization, see Section~\ref{sec:5}. As a result we achieve a good matching at $x=10^{-3}$ for all $Q^2$ of interest. However, this fit is not satisfactory, as we shall explain below.
\end{itemize}

\section{Description of the data  \label{sec:2}}
The LHCb collaboration have measured open \cc production at three different beam energies\footnote{Note that the measurements at 13 and 5 TeV have recently been corrected. We use these updated data.} 7, 13, 5 TeV \cite{LHCbcc7,LHCbcc13,LHCbcc5}. The data, $d^2\sigma/dp_{t,D}dy$, are presented for five $D$ meson rapidity intervals in the range $2<y<4.5$ and we use four transverse momentum bins covering the range  $1<p_{t,D}<5$ GeV. We fit to the data for $D^\pm, ~D^0, ~\bar{D}^0$ meson production.  

The $c\to D$ fragmentation functions $D(z)$ was taken from \cite{cac}, where they were determined from $e^+e^-$ annihilation data in the $\Upsilon(4S)$ region.
Actually the relative normalization of $c$ quark fragmentation to the different channels is only known from previous data to about 10$\%$ accuracy.  Therefore we allow, via a parameter $N^D$, for an additional renormalization of the $D^0,{\bar D}^0$ relative to the $D^\pm$ data.

The FONLL programme~\cite{FONLL} was used to calculate the open charm cross section at NLO.
 The running coupling, the charm mass (1.4 GeV), the quark and high $x$ gluon distributions are given by those found by the MMHT2014 NLO global fit \cite{Harland-Lang:2014zoa}.  We now describe, in turn, the three methods mentioned above, to determine the low $x$ gluon from the \cc LHCb data.

\section{Gluons at fixed scales with a simple parametrization  \label{sec:3}}
The data in each of the four $p_t$ intervals ($1-2,~2-3,~3-4,~4-5$ GeV) were fitted separately assuming a simple two-parameter form for the low  $x$ behaviour of the gluon
\be
xg(x)=N\left(\frac x{x_0}\right)^{-\lambda}
\label{eq:Nlam}
\ee
with $x=x_1$ and $x_0=10^{-5}$.  In this way we obtain gluons at four different scales, $\mu_F$.  Note that the value of $\mu_F$ is a bit larger than $\sqrt{p_{t,D}^2+m_D^2}$ since after fragmentation the transverse momentum of the $c$ quark is given by
\be
p_{t,c} ~=~p_{t,D}/z~ >~ p_{t,D}.
\ee
In fact $\mu_F= 2.0, ~2.9, ~3.9, ~4.9$ GeV for $p_{t,D}=1.5,~2.5,~3.5,~4.5$ GeV respectively. 

As mentioned above, we allowed an extra normalization parameter, $N^D$, between the $D^0,{\bar D}^0$ and the $D^\pm$ data. It was found to be
 $N^D\simeq 1.1$ in every case. In detail, this means we take the fractions 0.246 and 1.1(0.565) for the $D^\pm$ and $(D^0+{\bar D}^0)$ charm quark decay channels respectively, leaving 0.133 for the $D_s+\Lambda_c$ channels. The results are essentially unchanged if we set $N^D=1$, but then $\chi^2$ is a bit larger.
 
\begin{figure} [!htb]
\begin{center}
\vspace{-0.0cm}
\includegraphics[clip=true,trim=0.0cm .0cm 0.0cm 0.0cm,width=8.2cm]{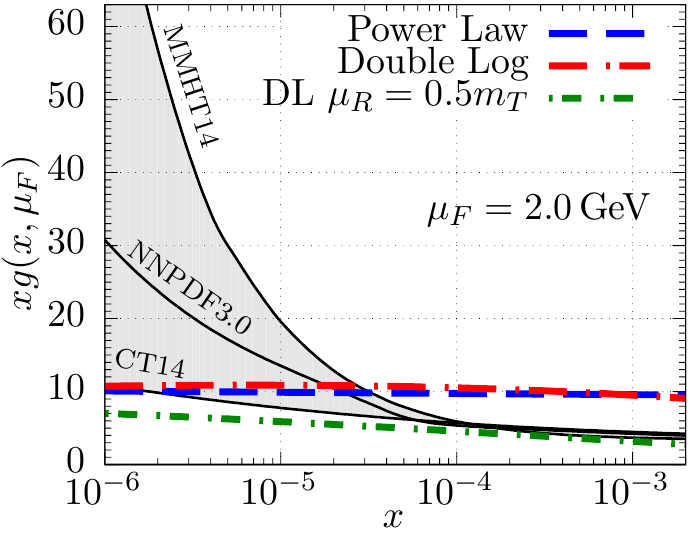} \hspace{.2cm}
\includegraphics[clip=true,trim=0.0cm .0cm 0.0cm 0.0cm,width=8.2cm]{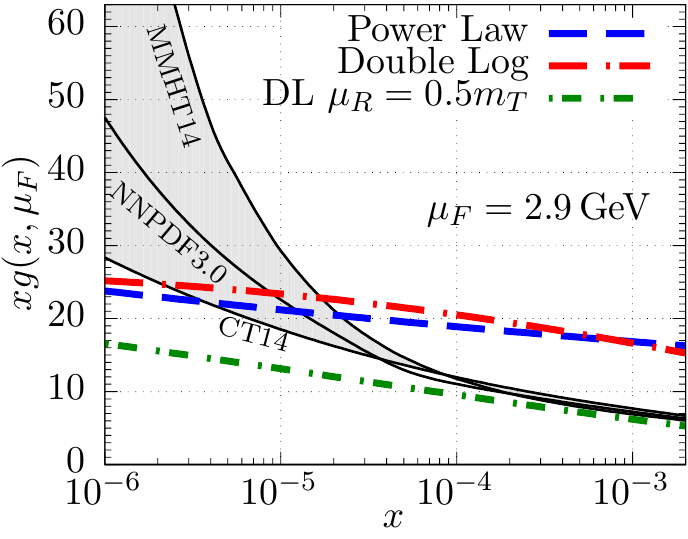}
\includegraphics[clip=true,trim=0.0cm .0cm 0.0cm 0.0cm,width=8.2cm]{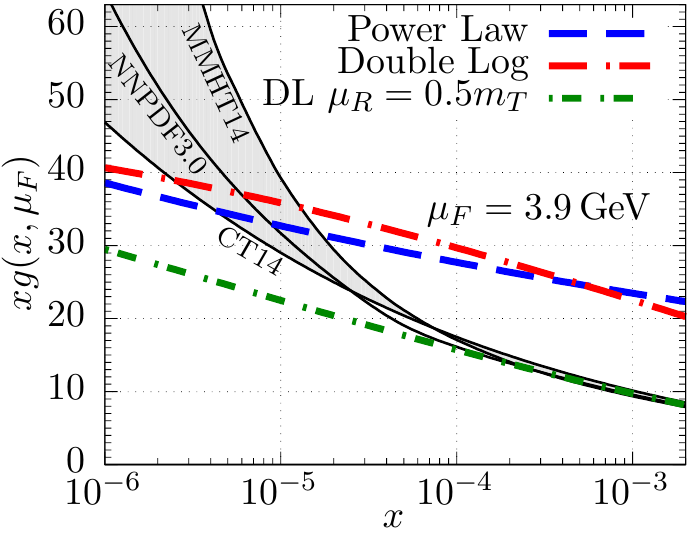} \hspace{.2cm}
\includegraphics[clip=true,trim=0.0cm .0cm 0.0cm 0.0cm,width=8.2cm]{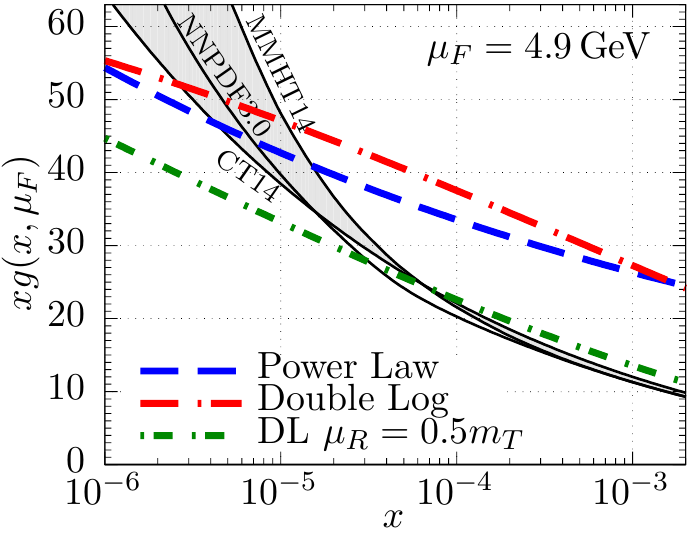}
\caption{\sf The four plots show the low $x$ behaviour of the gluon distribution in the four different $p_t$ intervals, obtained by three different fits to the LHCb open charm data \cite{LHCbcc7,LHCbcc13,LHCbcc5}. To be precise, the dashed curves are the gluon PDFs obtained by fitting to the data in each $p_{t,D}$ interval individually, see Sect.~\ref{sec:3},  whereas the long dash-dotted curves are the result of the `combined' fit
to the data in all four $p_{t,D}$ intervals simultaneously,
 see Sect.~\ref{sec:4}. The short dashed-dotted curves correspond to a combined fit with the further constraint that the gluon approximately matches the gluon of the global PDF analyses at $x=10^{-3}$ where it is well known, see Sect.~\ref{sec:5}. For reference, we also show the {\it central} values of the NLO gluon obtained by extrapolation using three different `global' PDF sets \cite{Ball:2014uwa}$-$\cite{Dulat:2015mca}; the uncertainties in the `global' gluons are much larger than that shown by the shaded bands in the above figures, see Fig~\ref{fig:g}.}
\label{fig:plots}
\end{center}
\end{figure}

 The normalization $N$ and the power $\lambda$ in (\ref{eq:Nlam}) and their uncertainties are obtained by fitting to the data in each $p_{t,D}$ interval using the MINUIT numerical minimization code~\cite{minuit,James:2004xla}.
The results of the four fits are presented in Table \ref{tab:1} and by the dashed curves in Fig. \ref{fig:plots}.
\begin{table}[hbt]
\label{tab:diff}
\begin{center}
\begin{tabular}{|c|c|c|c|c||c|c|c|}\hline
$p_{t,D}$ & $\mu_F$ & $N$ & $\lambda $ & $\chi^2_{\rm all} $ &  $\chi^2_5$ & $\chi^2_7 $ & $\chi^2_{13}$  \\

\hline
1.5&2.0& $9.9\pm0.4$&$0.01\pm 0.01$&51& 19 & 6 & 26 \\
  2.5& 2.9 & $21.2\pm 0.8$&$0.05\pm 0.01$&31& 11 & 6 & 13 \\
  3.5 & 3.9 & $32.7\pm 1.5$&$0.07\pm 0.01$&27& 7 & 12 & 8 \\
4.5 & 4.9& $42.7\pm 1.5$&$0.10\pm 0.01$&29& 4 & 14 & 11 \\
 \hline
\end{tabular}
\caption{\sf The parameters $N$ and $\lambda$ giving the low $x$ behaviour of the gluon distribution, $xg(x,\mu_F)=N(x/x_0)^{-\lambda}$ from individual fits to the LHCb open charm data \cite{LHCbcc7,LHCbcc13,LHCbcc5} in the four different $p_{t,D}$ intervals.  The scale $\mu_F$ and $p_{t,D}$ are given in GeV. The total $\chi^2$, $\chi^2_{\rm all}$, in each interval is shown, together with the contributions from the 5, 7 and 13 TeV data sets.}
\label{tab:1}
\end{center}
\end{table}

From the individual $\chi^2$ contributions in Table \ref{tab:1} we see hints of tension between the 5+13 TeV data on the one hand and the 7 TeV on the other hand. Surprisingly if the 5+13 TeV data are fitted without the 7 TeV data (and vice-versa) we find similar values of $N$ and $\lambda$, although the values of $\chi^2$ are reduced.

\section{Combined fit with a Double Log (DL) parametrization   \label{sec:4}}
As seen from the results of the simple fits shown in Table \ref{tab:1}, the gluon density increases strongly with scale, while the power of the $x$ behaviour has a weaker scale dependence. It is not evident whether such a behaviour is consistent with DGLAP evolution. Since in the low $x$ region the major effect comes from the DL contribution, it is reasonable to attempt to fit all four groups of data simultaneously with the formula
\be
xg(x,\mu^2)~=~N^{\rm DL}\left(\frac x{x_0}\right)^{-a}\left(\frac{\mu^2}{Q_0^2}\right)^b{\rm exp}\left[\sqrt{16(N_c/\beta_0){\rm ln}(1/x){\rm ln}(G)}\right]
\label{eq:DL}
\ee
with
\be
G=\frac{{\rm ln}(\mu^2/\Lambda^2_{\rm QCD})}{{\rm ln}(Q_0^2/\Lambda^2_{\rm QCD})}.
\label{eq:gPDF}
\ee
With three light quarks $(N_f=3)$ and $N_c=3$ we have $\beta_0=9$. The resummation of the leading double logarithmic terms $(\alpha_s{\rm ln}(1/x){\rm ln}(\mu^2))^n$ is written explicitly in the exponential, while the remaining single log terms are now parametrized by the powers $a$ and $b$. We take, in (\ref{eq:gPDF}), $\Lambda_{\rm QCD}=200$ MeV and $Q_0 =1$ GeV. As before, we set $x_0=10^{-5}$.   

Such an ansatz was used successfully to describe $J/\psi$ and $\Upsilon$ photoproduction data \cite{JMRT}. Moreover, it was checked that in the $x,~\mu_F^2$ region of interest this formula was consistent with NLO DGLAP evolution. So we fix the power $b$, which is responsible for the $\mu_F^2$ behaviour, to be the same as that found in the fit to $J/\psi$ photoproduction. Now  we are left to describe 120 LHCb data points with only two free parameters:  $N^{\rm DL}$ and $a$, plus the parameter $N^D$ introduced in Section~\ref{sec:3}.   

The parameters of the combined fit are presented in the first row of Table \ref{tab:2},
 and in Fig.~\ref{fig:plots} the results are compared with those of the simple fits. The low $x$ gluons obtained in the two fits are very similar. 
The DL description, which embodies NLO DGLAP evolution, should be more reliable than the results of the simple fits. Indeed, the dashed curves in Fig.~\ref{fig:fit} show that the `combined' fit gives an acceptable description of all the $D$ meson data in the interval $1<p_t<5$ GeV.

However, a problem is clearly evident. If the gluon PDF, determined from the LHCb data in the region of $x\sim 10^{-5}$, is extrapolated up to $x\sim 10^{-3}$ then it greatly exceeds the well known gluon densities determined by the global PDF analyses. Can the gluon density determined from the \cc data be reconciled with the global gluon PDF?

\section{A fit which matches to the `global' gluon at $x=10^{-3}$    \label{sec:5}}
Here we attempt to find a parametrization of the gluon PDF that fits the LHCb \cc data and which is consistent with the large $x$ gluon which is well known from the global PDF analyses \cite{Ball:2014uwa}$-$\cite{Dulat:2015mca}.  That is, we need to find a parametrization which reduces the extrapolated `\cc' gluon by more than a factor of two at $x\sim 10^{-3}$ for all $Q^2$ of interest.
We therefore choose  lower  
\vspace{-0cm}
\begin{figure} [H]
\vspace{-1cm}
\begin{center}
\includegraphics[scale=.98,trim=.9cm 0cm 0cm 10cm]{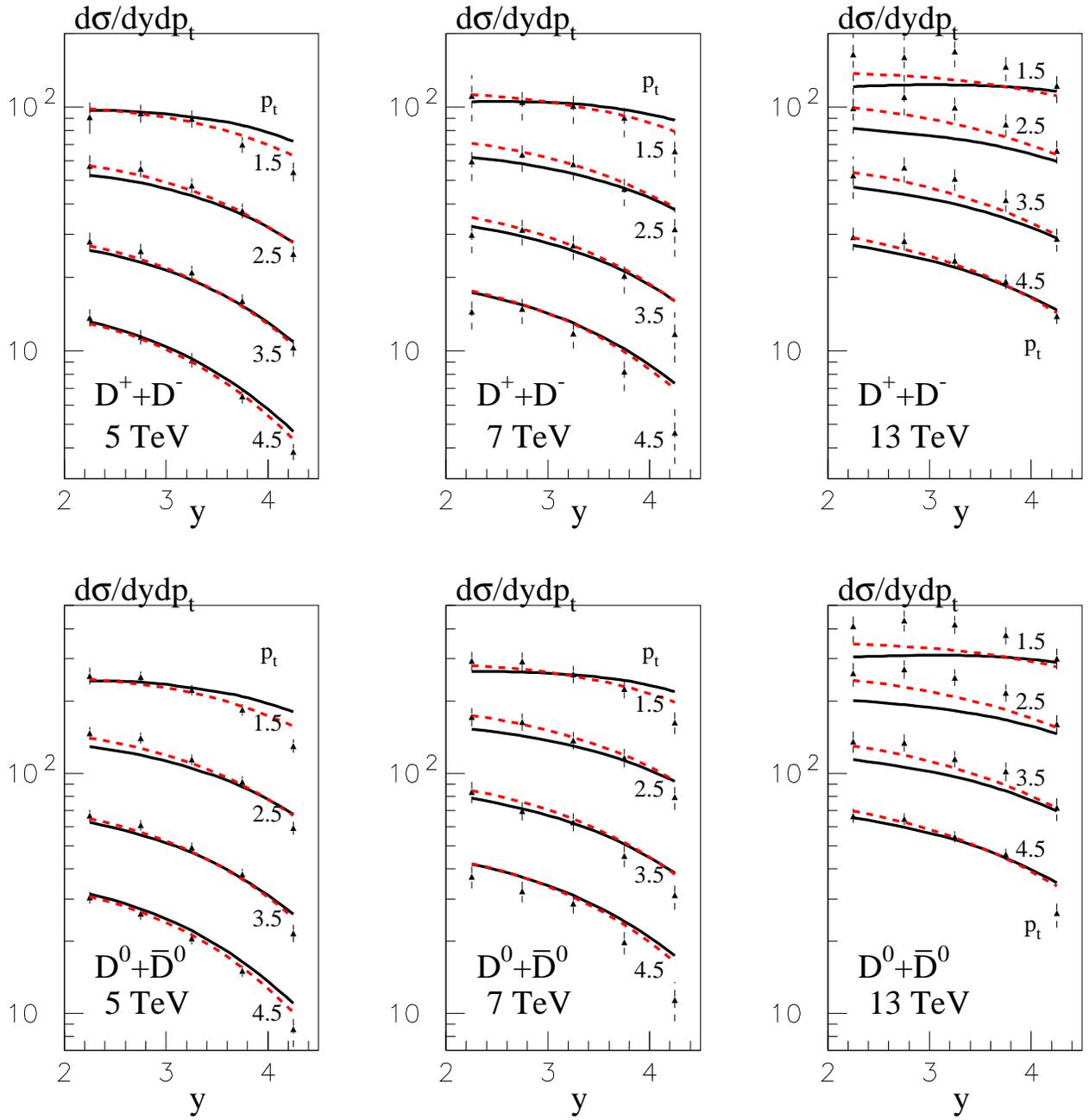}
\caption{\sf The rapidity dependence of the cross section of $D$ meson production (in $\mu$b/GeV) in four intervals of $p_t$ (in GeV). The data are from LHCb \cite{LHCbcc7,LHCbcc13,LHCbcc5}. The dashed curves are the fit with $Q_0$=1 GeV and $\mu_F=\mu_R=0.85m_T$ (Section~\ref{sec:4}), whereas the solid curves are the matching fit with $\mu_f=\mu_R =0.5m_T$ and $Q_0$=0.5 GeV (Section~\ref{sec:5}).}
\vspace{0.5cm}
\label{fig:fit}
\end{center}
\end{figure}
\begin{table}[hbt]
\label{tab:diff}
\begin{center}
\begin{tabular}{|c|c|c|c|c||c|c|c|}\hline
&  $N^{\rm DL}$ & $a$ &$b$ & $\chi^2$  &  $\chi^2_5$ & $\chi^2_7 $ & $\chi^2_{13}$  \\
\hline
Fit to \cc data (Sect.\ref{sec:4})&$0.13\pm 0.01$ & $-0.20\pm 0.01$ &-0.2 (fixed)&141 & 44 & 40 & 56\\
Fit\cite{JMRT} to $J/\psi$ data& $0.092\pm 0.009$& $-0.10\pm 0.01$&$ -0.2$ & & & &\\
  \hline
  \hline
  Fit to \cc data (Sect.\ref{sec:5}) & $0.0015\pm 0.0001$ & $-0.23\pm 0.01$ & $0.26 \pm 0.02 $ & 301 & 124 & 58 & 119 \\
  \hline
\end{tabular}
\caption{\sf The first row presents the values of the parameters $N^{\rm DL}$ and $a$ obtained in a fit to all the LHCb open charm data \cite{LHCbcc7,LHCbcc13,LHCbcc5} (120 data points in total) using the DL parametrization given by eqs. (\ref{eq:DL}) and (\ref{eq:gPDF}) of Section~\ref{sec:4}. We also show the contributions of the three data sets to the total $\chi^2=141$.  For comparison we show the parameters of the gluon obtained in a similar fit \cite{JMRT} to $J/\psi$ data. Note, however, that the gluons obtained in~\cite{JMRT} from $J/\psi$ data are not the $\overline{\mbox{MS}}$ gluons but correspond to those of the physical factorization scheme. The last row gives the values of the parameters of the fit described in Section~\ref{sec:5}; the much smaller value of $N^{DL}$ is compensated by a larger argument of the exponential factor in (\ref{eq:DL}) (due to $Q_0$ reducing from 1 to 0.5 GeV.)}
\label{tab:2}
\end{center}
\end{table}
factorization and renormalization scales\footnote{In the LO part of the contribution we still choose the `optimal' scale $\mu_F=0.85m_T$ which provides the resummation of the DL [(${\alpha_s N_C/\pi){\rm ln}(1/x){\rm ln}\mu_F^2}]^n$ terms, see~\cite{OMR}.  There still remains a dependence on the choice of the residual factorization scale $\mu_f$.} $\mu_f=\mu_R=0.5m_T$. The main effect comes from the change of renormalization scale, since the LO cross section is proportional to $\alpha_s^2(\mu_R^2)$. The gluon density does decrease, but it still strongly exceeds the global density at $x= 10^{-3}$. To overcome this problem we reduce the value of the parameter $Q_0$ in the DL form (\ref{eq:DL}). We take $Q_0=0.5$ GeV.  In this way we reach a satisfactory matching\footnote{Note that the matching is good for all $Q^2$ values. This is to be expected since the DL parametrization well approximates DGLAP evolution.} at $x= 10^{-3}$, at the price of considerably worsening the fit to the data, as can be seen by the solid lines in Fig.~\ref{fig:fit} and the $\chi^2$ values in the last row of Table~\ref{tab:2}.  Is this a problem of the fit to the data or it is a problem of the data themselves?

It appears that the rapidity dependence of the \cc cross section disagrees with the energy, $\sqrt{s}$, dependence of the data. Indeed, Fig.~\ref{fig:fit} shows that in order to describe the rapidity behaviour of the data, we need a curve that decrease faster with rapidity -- that is, a low $x$ gluon density which should increase {\it slower} with decreasing $x$ in comparison with that in the fit.  On the contrary, to obtain a larger $\sigma (13~{\rm TeV})/\sigma({5~\rm TeV})$ cross section ratio we need a low $x$ gluon which increases {\it faster} with decreasing $x$.

Another view of the same problem can be observed if we compare the data at larger beam energy $\sqrt{s_1}$ and large rapidity with data at a smaller beam energy $\sqrt{s_2}$~ and a smaller rapidity shifted by ln$(s_1/s_2)$.  In this case we deal with the same value of large $x_2$, so the ratio of cross sections should be equal to the ratio of the gluons at $x_1$ and $x_1(s_1/s_2)$.  The ratio of these cross sections is close to 1 for $D^0$, and is about $1.1-1.2$ for $D^+$, indicating that $\lambda\simeq 0-0.1$ in terms of (\ref{eq:Nlam}), in agreement with Table {\ref{tab:1}. This implies, that in this interval of $x$, the gluons are almost constant. However, in order to match these `flat' small $x$ gluons with the `global' gluons at $x=10^{-3}$, the `\cc' gluons have to decrease rapidly (by a factor of about 4) in the interval from $x=10^{-5}$ to $x=10^{-3}$; which corresponds to $\lambda\simeq 0.3$.  It is very unnatural to have such a rapid qualitative change in the $x$ behaviour of the gluon density.  No reasonable physical effect can generate this behaviour.

Note also, that contrary to \cc production, the same analysis of the ratio of $B$ meson cross sections indicates the behaviour of the low $x$ gluon is $xg\propto x^{-\lambda}$ with an unexpectedly large $\lambda>0.6$ for $x>10^{-4}$ \cite{Gauld}.

\section{Discussion}

Note that it looks quite unnatural to describe data with such a low renormalization scale as that in Section~\ref{sec:5}.  First, the idea of the collinear factorization approach was that the infrared behaviour of the amplitude is described by the incoming parton PDF, while the short range interaction is collected in the hard matrix element.  A low $\mu_R$ means that we include the running of the QCD coupling up to a large distance. Increasing the distance (as, for example, in the scales given in (\ref{eq:Acb}) for charm) means that we enter the region already described via the PDFs if $\mu_R <\mu_F$.  Technically if $\mu_R<\mu_F$ we have double counting of the loop insertion in the gluon propagator.  It is included in the DGLAP evolution of the PDF up to $\mu_F$ and the same loop is included in the renormalisation\footnote{See \cite{HKR} for a more detailed discussion.} of $\alpha_s$.  That is why we take $\mu_R=\mu_F$, and, in Section~\ref{sec:5}, $\mu_R=\mu_f$.

Thus we come back to the problem of matching at  larger $x$ with the gluons from the global analyses. To provide such a matching we need, in the interval of $x=0.5\cdot (10^{-4}-10^{-3})$, a larger value of $\lambda>0.3-0.4$. In general, the decrease of $\lambda$, with decreasing $x$, is expected. This is caused both, by stronger absorptive corrections at smaller $x$ and by the DL effects. The problem is that, already at $x=10^{-3}$, the  `global gluons' (gluons from global analyses) are already rather flat (see Fig.~\ref{fig:plots}, especially at a lowest scale ($p_t=1.5$ GeV).
For example, for $\mu_F=2.0$ (2.9) GeV the `global gluons' correspond to $\lambda \lapproxeq 0.17$ (0.23) at $x=10^{-3}$ and there is no reason for $\lambda$ to increase at $x<10^{-3}$.

\subsection{Alternative parametrization}
In the present paper we have used the DGLAP form of the gluon distribution in order to compare the low $x$ gluon PDF obtained from the analysis of the LHCb open charm data, with that given by the parton global analyses. On the other hand, at such small values of $x$  the QCD dynamics
may be better described by the BFKL equation. The solution of BFKL equation 
leads to a power-like $x^{-\lambda}$ behaviour (analogous to used in eq.~(\ref{eq:Nlam}) above) with a prefactor which weakly depends on $x$. That is, the analysis in Sect.~\ref{sec:3} can be considered
as a BFKL-based description of the data. Since the analysis was done for each $p_t$ bin individually, we do not have to worry about the form of the scale (or $p_t$) dependence of the gluon parametrization.

If  we were to include the prefactor, and write the asymptotic BFKL amplitude as
\begin{equation}
A(x)~=~\frac{x^{-\lambda}}{\sqrt{\ln(1/x)}}~\propto xg(x)
\end{equation}
then the matching with the high $x$ global gluons would be even worse. Indeed,
the effective power 
\begin{equation}
\lambda_{\rm eff}= - \frac{d \ln(A(x))}{d \ln(x)} = \lambda-1/(2\ln(1/x))
\end{equation} would decrease with increasing $x$ and the extrapolation of our solution to larger $x$ would give even larger gluons than those shown in Fig.\ref{fig:plots}.

However, the low value of $\lambda\sim 0.01-0.1$ indicates that in our relatively low-$p_t$ domain we do not deal with a pure  BFKL amplitude (that is a single QCD Pomeron). The expected BFKL intercept should be larger. At NLL
level the value of $\lambda=0.2-0.3$ and weakly depends on the 
scale, see, for example~\cite{bfkl1,bfkl2}.

It might be thought that
the low values of $\lambda$, that we obtain in Table 1, may be explained by a possible saturation effect; since, as $x$ decreases, the growth of absorptive corrections damps down the rise of the low-$x$ gluons. As is seen from Fig.\ref{fig:plots}, the extrapolation of the conventional `global gluons' leads, at $x\sim 10^{-5}$, to gluon densities  which are consistent with our gluons (needed to describe the LHCb open charm data), and {\em if} then the value $\lambda$ will be strongly diminished
 due to the saturation effect, this could provide reasonable matching with the larger $x$
 `global gluons'.  However, this explanation does not work in practice.

The problem is the $p_t$ dependence. For a smaller value of $p_t$ the cross section of an additional interaction is larger, and consequentially  the absorptive corrections are stronger. So, for example, 
in the $p_t=4.5$ GeV bin ($\mu_F=4.9$ GeV) we expect a much weaker effect than that
for $p_t=1.5$ GeV ($\mu_F=2$ GeV). Moreover, for larger $p_t$, we sample larger $x$. On the other hand, we see from Fig.~\ref{fig:plots} that
at, say, $x=10^{-4}$ and $\mu_F=3.9$ or 4.9 GeV, the density of `global gluons' is
already about twice smaller than that from our analysis (needed to describe the data) and including the saturation effect we will only diminish it further, and make the discrepancy larger.
 That is, while there is a chance to get more or less a satisfactory description of  the lowest $p_t$ bin, by tuning the parameters of the absorptive (saturation) effect, we will surely fail to describe the data at larger values of $p_t$.

Besides this, we have to recall, that {\em no} saturation effects were observed in
the analysis of data for exclusive $J/\psi$ production  at the LHC. These data, which probe more or less the same kinematic region\footnote{The scale and $x$ values are just a bit smaller.},  are well described by the DL parametrization of eq.~\ref{eq:DL} (see e.g.~\cite{JMRT}).

\subsection{Consistency of heavy quark data}
\begin{figure} [!htb]
\begin{center}
\vspace{-0.0cm}
\includegraphics[clip=true,trim=0.0cm .0cm 0.0cm 0.0cm,width=8.2cm]{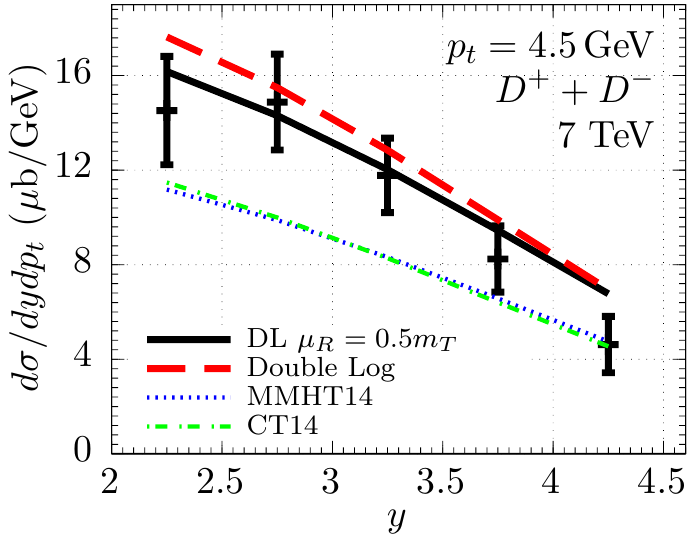} \hspace{.2cm}
\includegraphics[clip=true,trim=0.0cm .0cm 0.0cm 0.0cm,width=8.2cm]{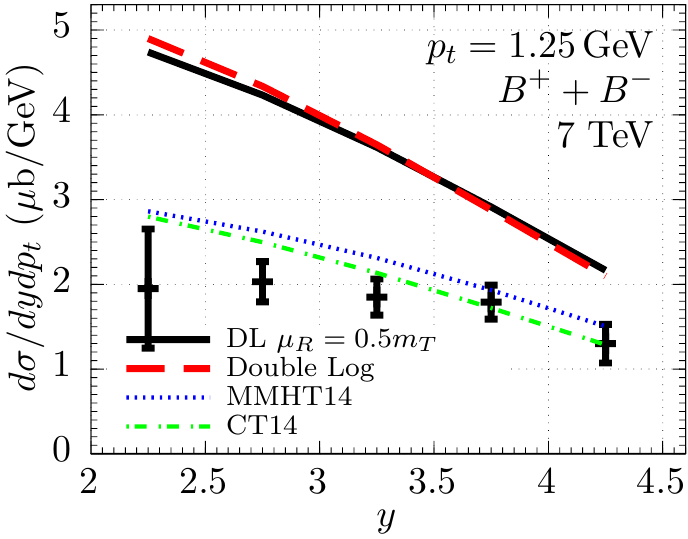}
\caption{\sf Plots of $d\sigma/dydp_t$ versus $y$ for $D$ and $B$ meson production at $p_t$ values which correspond to approximately the same scales and the same values of $x$. The bold dashed and solid curves are obtained from the gluons fitted to the $D$-meson data as described in Sections \ref{sec:4} and \ref{sec:5} respectively.}
\label{fig:DB}
\end{center}
\end{figure}

As discussed in \cite{Gauld}, the disagreement between the experimentally observed charm/bottom cross sections and the QCD predictions may be caused by a deficiency in calculating the detector efficiencies in the different $y$ intervals. Recall that the experimental rapidity behaviour for the $D$-meson cross section leads to a very flat low $x$ gluon PDF, $xg$, see the low value of $\lambda$ in Table~\ref{tab:1}. When this gluon is extrapolated to larger $x$ we have a large mismatch with the well known gluons of the global PDF analyses.  
On the other hand, to describe
the $B$-meson rapidity distribution~\cite{LHCbB1} (which is rather flat in
$y$) we need much steeper gluons.  Fig.~\ref{fig:DB} shows an example of comparing $D$ and $B$ meson production. The figure compares the QCD
predictions calculated with different sets of gluon PDFs for the 7 TeV
data taken at $p_t=4.5$ GeV for $D$-meson production (the left plot) with a corresponding plot of $B$-meson data at $p_t=1.25$ GeV. These two plots correspond to practically the same scale (the same value of
$m_T$) and the same $x$ interval. We see that the gluons found in the present paper, which describe the $D$-meson cross
sections, strongly overestimate $B$-meson production. Moreover, the
$B$-meson data ask for even steeper gluons than those given by the central
values of the gluon PDFs of the CT14 and MMHT2014 global parton analyses.
Indeed, to reproduce such a flat rapidity dependence of the $B$-meson cross section shown in Fig.~\ref{fig:DB}, the gluons should increase faster with decreasing $x$.}

Note also that even using the low renormalization scale $\mu_R=0.5m_T$, and the correspondingly smaller gluons
which at $x$=0.001 match those from the global analyses, we 
overestimate the $D$-meson cross section measured at $\sqrt s=1.96$ TeV by the
CDF collaboration~\cite{CDF} in just the $x\sim 0.001$ region. For $p_t=2$
GeV we find 53.8 $\mu$b/GeV as compared to the CDF result of $32.7\pm 6.5\pm 4.2$
$\mu$b/GeV. The predicted cross section is too large -- not due to the new (larger) gluon PDFs, but due to the large value of $\alpha_s(\mu_R)$ taken at the low scale $\mu_R=0.5m_T$.

All these facts indicate an inconsistency in the data.

Another possibility is that at these relatively low scales (that is, large distances) relevant for charm production we feel non-perturbative effects. Note that a large discrepancy is observed in Fig.~\ref{fig:fit} for the smallest $p_t$ bin. There could be the non-perturbative production of a \cc pair or non-perturbative effects in the $c\to D$ fragmentation. Unlike the $e^+e^-$ case, for inclusive $pp$ processes the $c$ quark is surrounded by a large number of light quarks from the underlying event. Then, besides fragmentation, the $D$ meson may be formed by the fusion of a charm and a light quark.  To investigate the latter case it would be interesting to see the dependence of $D$ meson production in events with different multiplicities.

However a similar problem is observed for $B$ meson production (see (\ref{eq:Acb}) and \cite{R2, Gauld}) where we deal with a larger scale and so the non-perturbative effects will be heavily suppressed. Therefore, in order to isolate the role of non-perturbative effects, it would be very interesting to see more precise and more differential data, especially for low $p_t$ $B$ meson production.

Recall that in spite of the large mass $m_b$, there is the possibility to probe low factorization scales ($\mu_F<m_b)$ by observing {\it both} the $B$ and ${\bar B}$ mesons and selecting the events with small acoplanarity, see the discussion in Section 5 of \cite{OMR}.

\section*{Acknowledgements}
MGR thanks the IPPP at the University of Durham for hospitality. This work was supported by the RSCF grant 14-22-00281 for MGR and by Capes, Fapesc, INCT-FNA (464898/2014-5), and CNPq (Brazil) for EGdO.



\thebibliography{}

\bibitem{LHCbcc7} R. Aaij {\it et al.} [LHCb Collaboration], Nucl. Phys. {\bf B871} (2013) 1.   

\bibitem{LHCbcc13}  R.~Aaij {\it et al.} [LHCb Collaboration], JHEP {\bf 1603} (2016), 159. Erratum: JHEP {\bf 1609} (2016), 013; JHEP {\bf 1705} (2017), 074.

\bibitem{LHCbcc5} R.~Aaij {\it et al.} [LHCb Collaboration], arXiv:1610.02230v5.

\bibitem{LHCbB1} R.Aaij {\it et al.} [LHCb Collaboration], JHEP {\bf 1308} (2013) 117 [arXiv:1306.3663].

\bibitem{LHCbB} R. Aaij {\it et al.} [LHCb Collaboration], Phys. Rev. Lett. {\bf 118} (2017) 052002 [arXiv:1612.05140].

\bibitem{R1} R. Gauld, J. Rojo, L. Rottoli and J. Talbert, JHEP {\bf 1511} (2015) 009. [arXiv:1506.08025].

\bibitem{R2} O. Zenaiev {\it et al.}, Eur. Phys. J. {\bf C75} (2015) 396 [arXiv:1503.04581].

\bibitem{R3} M. Cacciari, M.L. Mangano and P. Nason, Eur. Phys. J. {\bf C75} (2015) 610 [arXiv:1507.06197].
 
\bibitem{GR}  R. Gauld and J. Rojo, Phys. Rev. Lett. {\bf 118} (2017) 072001  [arXiv:1610.09373].

\bibitem{Gauld} R. Gauld, JHEP {\bf 1705} (2017) 084, [arXiv:1703.03636 [hep-ph]]. 
%
\bibitem{Ball:2014uwa}
  R.D.\ Ball {\it et al.}  [NNPDF Collaboration],
  JHEP {\bf 1504} (2015) 040.
%
\bibitem{Harland-Lang:2014zoa}
  L.A.\ Harland-Lang, A.D.\ Martin, P.\ Motylinski and R.S.\ Thorne,
  Eur.\ Phys.\ J.\ {\bf C75} (2015) 5, 204.
%
\bibitem{Dulat:2015mca}
  S.\ Dulat, T.J.\ Hou, J.\ Gao, M.\ Guzzi, J.\ Huston, P.\ Nadolsky,
  J.\ Pumplin, C.\ Schmidt, D. Stump and C.P. Yuan, Phys. Rev. {\bf D93} (2016) 033006 [{\tt arXiv:1506.07443}].

\bibitem{Buckley:2014ana} 
A.~Buckley, J.~Ferrando, S.~Lloyd, K.~Nordström, B.~Page, M.~Rüfenacht, M.~Schönherr and G.~Watt,
Eur.\ Phys.\ J.\ C {\bf 75}, 132 (2015)
doi:10.1140/epjc/s10052-015-3318-8
[arXiv:1412.7420 [hep-ph]].
 
\bibitem{OMR}E.G. de Oliveira, A.D. Martin and M.G. Ryskin, Eur. Phys. J. {\bf C77} (2017) 182 [{\tt arXiv:1610.06034}].
 
 \bibitem{cac} M. Cacciari, P. Nason and C. Oleari, JHEP {\bf 0604} (2006) 006 [hep-ph/0510032]. 

\bibitem{FONLL} S. Forte, E. Laenen, P. Nason and J. Rojo, Nucl. Phys. {\bf 834} (2010) 116.

\bibitem{minuit} F. James and M. Roos, Compt. Phys. Commun. {\bf 10}, (1975) 343.

\bibitem{James:2004xla}
F.~James and M.~Winkler,
``MINUIT User's Guide.''

\bibitem{JMRT} S.P. Jones, A.D. Martin, M.G. Ryskin and T. Teubner, J. Phys. G {\bf 44} (2017) 03LT01 [arXiv:1611.03711].

\bibitem{HKR} L.A. Harland-Lang, V.A. Khoze and M.G. Ryskin,
 Phys. Lett. {\bf B761} (2016) 20
 [arXiv:1605.04935]. 
 
 \bibitem{bfkl1} S.J. Brodsky, V.S. Fadin, V.T.
Kim, L.N. Lipatov, G.B. Pivovarov, JETP Lett. 70 (1999)
[hep-ph/9901229].

\bibitem{bfkl2} G.~P.~Salam,
  Acta Phys.\ Polon.\ B {\bf 30}, 3679 (1999)
  [hep-ph/9910492];
JHEP {\bf 9807} (1998) 019,
  [hep-ph/9806482].

\bibitem{CDF} T.A. Aaltonen {\it et al.} [CDF Collaboration], Phys. Rev. {\bf D95} (2017) 092006 [arXiv:1610.08989].
\end{document}